\documentclass[%
 reprint,
superscriptaddress,
 amsmath,amssymb,
 aps,
pra,
]{revtex4-2}

\usepackage{graphicx}
\usepackage{dcolumn}
\usepackage{bm}

\begin{document}


\title{Unbalanced droplets of one-dimensional mixtures of fermions}

\author{M. C. Gordillo}
\email{cgorbar@upo.es}
\affiliation{Departamento de Sistemas F\'{\i}sicos, Qu\'{\i}micos y Naturales, Universidad Pablo de
Olavide, Carretera de Utrera km 1, E-41013 Sevilla, Spain}
\affiliation{Instituto Carlos I de F\'{\i}sica Te\' orica y Computacional, Universidad de Granada, E-18071 Granada, Spain.}

\date{\today}

\begin{abstract}
By means of a diffusion Monte Carlo technique,  we study one-dimensional unbalanced mixtures of fermionic Ytterbium atoms ($^{173}$Yb, $^{171}$Yb). 
This means clusters in which the total number of $^{173}$Yb particles is different from the sum of all the atoms belonging to the $^{171}$Yb isotope.  Our aim will be to check the possibility of having self-bound arrangements
beyond the balanced compositions reported in previous literature
rather than exploring all the situations in which that could be
possible.
In that vein, we focused mainly  on mixtures in which the atoms belonging to one isotope are spin-polarized, while the spins of the particles in the other isotope are evenly distributed in two 
sets.  
What we found was that, even tough self-bound droplets are possible for different compositions,  the most stable ones are clusters with a slight excess of attractively interacting $^{171}$Yb particles 
with different spins with respect to the number of spin-polarized $^{173}$Yb atoms. Clusters in which the number of repulsively interacting unequal-spin $^{173}$Yb atoms are in excess with respect to the spin-polarized $^{171}$Yb particles have a very narrow stability range.   
\end{abstract}

\maketitle

\section{Introduction}

A droplet can be defined as a (relatively) small cluster of particles that stick together without collapse or evaporation during a reasonable long period of time.  To be considered self-bound,  a droplet has to be 
stable without the intervention of a external confining potential.
When several species are present, the outcome of the mixing  would depend on factors such as the bosonic or
fermionic nature of the  particles,  their repulsive or attractive interactions and the dimensionality of the 
system \cite{rep2020,frontiers2021,few1d}.  
The seminal work by Petrov \cite{petrov2015},  opened the field for ultracold Bose-Bose droplets  \cite{rep2020,frontiers2021,cabrera,soliton,freedrop,dipolar1,errico,guo,
parisi,petrov2,njp2020},  but other possibilities  \cite{sal1,njp2016,scipost2019,njp2019,peacock,yoprr,yopral,tononi} are also viable.

Within this context, the consideration of one-dimensional (1D) systems offers the advantage 
of the suppression of three-body losses with respect to their three-dimensional counterparts \cite{grelosses,few1d}.   In addition,  if the 
interactions between species are the right ones, we can have 
self-bound Bose-Bose \cite{parisi,petrov2,njp2020},  Bose-Fermi \cite{sal1,njp2019},   and 
Fermi-Fermi 1D droplets \cite{yoprr,yopral}.  
In this last case, 
when the number of spin-ups equals the number of spin-downs, 
we end up with a set of composite bosons (or "molecules")
with an effective repulsion between them \cite{shlyapnikov,jordibcs} that makes 
self-bound balanced 1D clusters of equal-mass fermions impossible \cite{1dfermion,yoprr}. 

Fortunately,  we can 
circumvent that limitation by considering at least three different types of fermionic species \cite{yoprr,yopral}.  This has been proved to work  in small clusters of Ytterbium ($^{173}$Yb and $^{171}$Yb) with attractive short-range interactions between atoms of different isotopes.  Since 
atoms belonging to the same isotope can have different spins, 
not all the $^{173}$Yb-$^{171}$Yb molecules are equal: we have as many types of 
composite bosons as possible spin pairings.  Then, the relaxation of the     
Pauli avoidance between different kinds of molecules 
allows the attractive interaction between them to kick in,  producing self-bound droplets.  

What all previous cases have in common is that they consider only {\em balanced} clusters, i.e., 
ensembles in which the number of atoms of both species (or isotopes)  is equal to each other. 
Very recently, that constraint has been removed in some studies of Bose-Bose clusters
modeled by Gross-Pitaevskii equations 
\cite{dropletboson1,dropletboson2,dropletboson3,dropletboson4} or within the framework of the
discrete 1D Bose-Hubbard model \cite{scipost2024}. 
All those works agreed in the existence of stable unbalanced clusters of particles,
at least within certain values of the parameters defining the droplets.  
This opens the door to the study of differences in the behavior of those clusters with 
respect to the case of the balanced ones and to establish their stability ranges.   

In this work, we will expand this new avenue by considering continuous 1D systems of
self-bound unbalanced {\em fermions}.  To do so, we will deal 
with 
$^{173}$Yb-$^{171}$Yb mixtures in which $N_{173} \ne N_{171}$ ($N_{173}$ and $N_{171}$ being their
respective number of particles).
We aim only to check whether those unequal particle clusters exist,  rather than exhaustively testing all the situations in which we can find unbalanced self-bound setups. 
This is the reason why,  to simplify things, 
we will consider mainly situations in which one of the components (either $^{173}$Yb and $^{171}$Yb) is spin-polarized while the atoms of the other are equally split into two unequal spin sets.  
In this, we are guided by the results of balanced clusters, in which
all compositions including more than three types of spins or in which the numbers of atoms with the same spin are not the same, have
qualitatively similar behaviors when the total number of particles is fixed \cite{yoprr,yopral}.  

\section{Method}

Following the previous literature,  the 1D clusters in this work will be described by the following Hamiltonian \cite{su6su2,su6su2ol,yoprr,yopral}: 
\begin{eqnarray}\label{hamiltonian}
 H=\sum_{i=1}^{N_p}\frac{-\hbar^2}{2m}\nabla_i^2 + g_{1D}^{173-171} \sum_{i=1}^{N_{173}} \sum_{j=1}^{N_{171}} \delta(x^{173}_i - x^{171}_j)
 \nonumber \\
+ g_{1D}^{173-173} \sum_{b>a} \sum_{i=1}^{n_{173,a}} \sum_{j=1}^{n_{173,b}} \delta(x^{173}_{a,i} - x^{173}_{b,j})
 \nonumber \\
+ g_{1D}^{171-171} \sum_{b>a} \sum_{i=1}^{n_{171,a}} \sum_{j=1}^{n_{171,b}} \delta(x^{171}_{a,i} - x^{171}_{b,j}),
\end{eqnarray}
where $N_p$ is the total number of fermions, with $N_p$ = $N_{173}$+$N_{171}$, and will be in the range 24-36.  As indicated above, 
$N_{173} \ne N_{171}$.  Eq. \ref{hamiltonian} considers only interactions in which the atoms in the pair are different from each other, since Pauli's exclusion principle takes care of the avoidance between identical fermions. 
$m$ is the mass of the atoms, described by a single parameter as in the previous literature \cite{yoprr,yopral,su6su2,su6su2ol}.  
This is expected to be a reasonable approximation, since the 
mass difference between isotopes is around 1 \%. 
$n_{173,{ab}}$ and $n_{171,{ab}}$ are the number of atoms with spins $a$ and $b$. The $g_{1D}$ parameters depend on the 1D-scattering lengths, $a_{1D}$, via $g_{1D}^{\alpha,\beta} = -2 \hbar^2/m a_{1D}(\alpha,\beta)$.  $a_{1D}$ is defined by \cite{Olshanii}:
\begin{equation} \label{a1D}
a_{1D}(\alpha,\beta) = -\frac{\sigma_{\perp}^2}{a_{3D}(\alpha,\beta)} \left( 1 - A \frac{a_{3D}(\alpha,\beta)}{\sigma_{\perp}} \right),
\end{equation}
with A=1.0326 and $(\alpha,\beta)$ = $(173,171)$. $\sigma_{\perp} = \sqrt{\hbar/m \omega_{\perp}}$ is the  oscillator length  in the transverse direction, depending on the perpendicular confinement frequency, $\omega_{\perp}$, taken to be in the range 2$\pi\times$50-100 kHz, tight enough to produce a quasi-one dimensional system. $a_{3D}(\alpha,\beta)$ are the three-dimensional experimental scattering lengths between isotopes, taken from Ref.  \onlinecite{pra77}
i.e., 10.55 nm ($^{173}$Yb-$^{173}$Yb), -0.15 nm ($^ {171}$Yb-$^{171}$Yb) and -30.6 nm ($^{171}$Yb-$^{173}$Yb), where the minus signs mean attractive interactions.  Those scattering lengths cannot be changed via magnetic Feshbach resonances due to the closed-shell electronic structure of Yb atoms.  The use of other kinds of Feshbach resonances is, at best,  problematic \cite{cazalilla}, and to our knowledge,  it has only been studied for setups containing a single Yb isotope \cite{Yb174,Yb171,Yb173}.
However,  the values of $g_{1D}^{\alpha,\beta}$ can be changed by 
modifying the external confinement,  $\sigma_{\perp}$,  via 
Eq. \ref{a1D}.  That variation could also, in principle,  change
the nature of the interactions, determined by the sign of $a_{1D}$,
through a confining-induced resonance.  To do so, $A a_{3D}/\sigma_{\perp}$ should be greater than 1, something 
that does not happen for any of the $a_{3D}$ values in the considered range of frequencies.  This means that the
nature of the interactions between isotopes is fixed by the sign of those $a_{3D}$'s: atractive for the $^{173}$Yb-$^{171}$Yb and  $^{171}$Yb-$^{171}$Yb pairs and repulsive in the   $^{173}$Yb-$^{173}$Yb case.   
                 
To solve the Schr\"odinger equation derived from the continuous Hamiltonian in Eq. \ref{hamiltonian}, we used the  fixed-node diffusion Monte Carlo 
(FN-DMC) algorithm, that gives us the exact ground state of a 1D system of fermions \cite{hammond,ceperley}, without 
resorting to any mean field treatment.  We start from an initial approximation to the exact wavefunction, the so-called trial function.  
We used:
\begin{eqnarray} \label{defa}
\Phi(x_1,\cdots,x_{N_p}) = \nonumber \\
 \mathcal{A}(x_1^{173},x_2^{173},\cdots,x_{N_{173}}^{N_{173}}, x_1^{171}, x_2^{171},\cdots,x_{N_{171}}^{N_{171}})
 \nonumber \\
\prod_{b>a}
\prod_{i=1}^{n_{173,a}} \prod_{j=1}^{n_{173,b}}  \frac{\psi (x_{a,i}^{173}-x_{b,j}^{173})}{(x_{a,i}^{173}-x_{b,j}^{173})} \nonumber \\
\prod_{b>a}
\prod_{i=1}^{n_{171,a}} \prod_{j=1}^{n_{171,b}}  \frac{\psi (x_{a,i}^{171}-x_{b,j}^{171})}{(x_{a,i}^{171}-x_{b,j}^{171})},
\end{eqnarray}
where $\mathcal{A}(x_1^{173},x_2^{173},\cdots,x_{N_{173}}^{N_{173}}, x_1^{171}, x_2^{171},\cdots,x_{N_{171}}^{N_{171}})$ is the determinant of a square matrix that depends on the coordinates of 
all the particles in the system.  The dimension of that square matrix will be $N_{max} \times  N_{max}$,  $N_{max}$ being the maximum 
value between $N_{173}$ and $N_{171}$. To build that matrix,  we followed the prescription used 
in Ref. \onlinecite{pandaripande} for unbalanced sets of three-dimensional fermions.  If,  for the sake of the argument,  we consider $N_{max} = N_{171} > N_{173}$, we will have a 
$N_{171} \times N_{171}$ matrix.  Bearing in mind that 
the solutions of the Sch\"odinger equation for a pair of 1D-particles
interacting via an attractive delta potential can be written as \cite{griffiths}: 
\begin{equation} \label{solution1D}
\phi(|x_i^{173}-x_{j'}^{171}|) = \exp \left[-\frac{|g_{1D}^{173,171}|}{2} |x_i^{173}-x_{j'}^{171}| \right],
\end{equation}
then, we have that the first $N_{173}$ rows of that $N_{171} \times N_{171}$ matrix 
are of the form:
\begin{equation}
\phi (|x_i^{173}-x_{1'}^{171}|),   \phi (|x_i^{173}-x_{2'}^{171}|),  \cdots,  \phi (|x_i^{173}-x_{N_{171'}}^{171}|),
\end{equation}
with $i$ in the range $i$=1,$\cdots$,$N_{173}$.  The remaining $N_{171}-N_{173}$ rows have to include 
functions that depend exclusively on the coordinates of all the atoms of the  $^{171}$Yb isotope.  Following
Ref. \onlinecite{sola},  we considered single-particle orbitals similar to the ones we would have in a Vandermonde
matrix \cite{girardeau}. This means that the complete $N_{171} \times N_{171}$ determinant can be written as:
\begin{center}
$\left|
\begin{array}{cccc}
x_{1'}^{N_{171}-N_{173}-1} & x_{2'}^{N_{171}-N_{173}-1} & \cdots & x_{N_{171}'}^{N_{171}-N_{173}-1} \\
\cdots & \cdots & \cdots & \cdots   \\
x_{1'}^2 & x_{2'}^2 & \cdots & x_{N_{171}'}^2 \\
x_{1'} & x_{2'} & \cdots & x_{N_{171}'} \\
1 &   1 &  \cdots & 1 \\
\phi (r_{N_{173},1'}) &  \phi (r_{N_{173},2'}) & \cdots & \phi (r_{N_{173},N_{171}'}) \\
 \cdots & \cdots & \cdots & \cdots   \\
\phi (r_{2,1'}) &  \phi (r_{2,2'}) & \cdots & \phi (r_{2,N_{171}'}) \\
\phi (r_{1,1'}) &  \phi (r_{1,2'}) & \cdots & \phi (r_{1,N_{171}'}) 
\end{array}
\right|$ \label{row}
\end{center}
with $r_{i,j'}$ = $|x_i^{173}-x_{j'}^{171}|$. This form does not include any confining wavefuncion, as it is typically the case when an harmonic potential is included in the Hamiltonian, \cite{girardeau} and generalizes the used in previous literature  for balanced fermion clusters. 

The terms $(x_{a,i}^{\alpha}-x_{b,j}^{\alpha})$ in the denominator of Eq. \ref{defa} correct the spurious nodes 
between atoms of the same isotope with different spins (see Refs. \onlinecite{su6su2,yoprr} for further details). 
In Eq. \ref{defa}, $\psi (x_{a,i}^{\alpha}-x_{b,j}^{\alpha})$'s are Jastrow functions that introduce the correlations
between pairs of particles of the same isotope belonging to different spin species $a,b$.
For the repulsively interacting $^{173}$Yb-$^{173}$Yb pair,  we have \cite{gregoritesis}:
\begin{equation}\label{delta}
\psi (x_{a,i}^{173}-x_{b,j}^{173}) = \cos(k[|x_{a,i}^{173}-x_{b,j}^{173}|-R_m])
\end{equation}
when the distance between atoms, $|x_{a,i}^{173}-x_{b,j}^{173}|$, was smaller than a variationally obtained parameter, $R_m$, and 1 otherwise. $k$ was obtained by solving the transcendental equation $k a_{1D}(173,173)\tan(k R_m)=1$. When the pair of particles of the same isotope attract each other, as in the $^{171}$Yb-$^{171}$Yb case, the Jastrow has
the form of Eq.  \ref{solution1D} \cite{gregoritesis,su6su2},
but with a different value of the defining constant, $g_{1D}^{171,171}$. 

\section{Results}

As shown in previous literature \cite{1dfermion,yoprr}, when $N_{173}$=$N_{171}$ and the atoms belonging to both isotopes are spin-polarized, we have a set of composite bosons with an effective repulsion between them due to a double Pauli avoidance that precludes the formation of self-bound clusters. One of the tell-tale signals of this behavior is that
the energy per particle is exactly $E_b/2$ =  -$(g_{1D}^{173,171})^2/(8 \hbar \omega_{\perp} \sigma_{\perp})$,
with $E_b$ the binding energy of a pair of particles interacting attractively via a 1D delta potential \cite{griffiths}.
On the other hand,  when $N_{171}>N_{173}$ (again for the sake of the argument) what we have is that the total energy of the system is 
$N_ {173} E_b$ and the system is again unbound.  As stated above,  we have neglected the mass difference between isotopes ($\sim$ 1 \%) even tough that difference,  if large enough, could 
induce the existence of self-bould boson \cite{paper2009} and 
fermion \cite{paper2009,givois} arrangements.

\begin{figure}
\begin{center}
\includegraphics[width=0.8\linewidth]{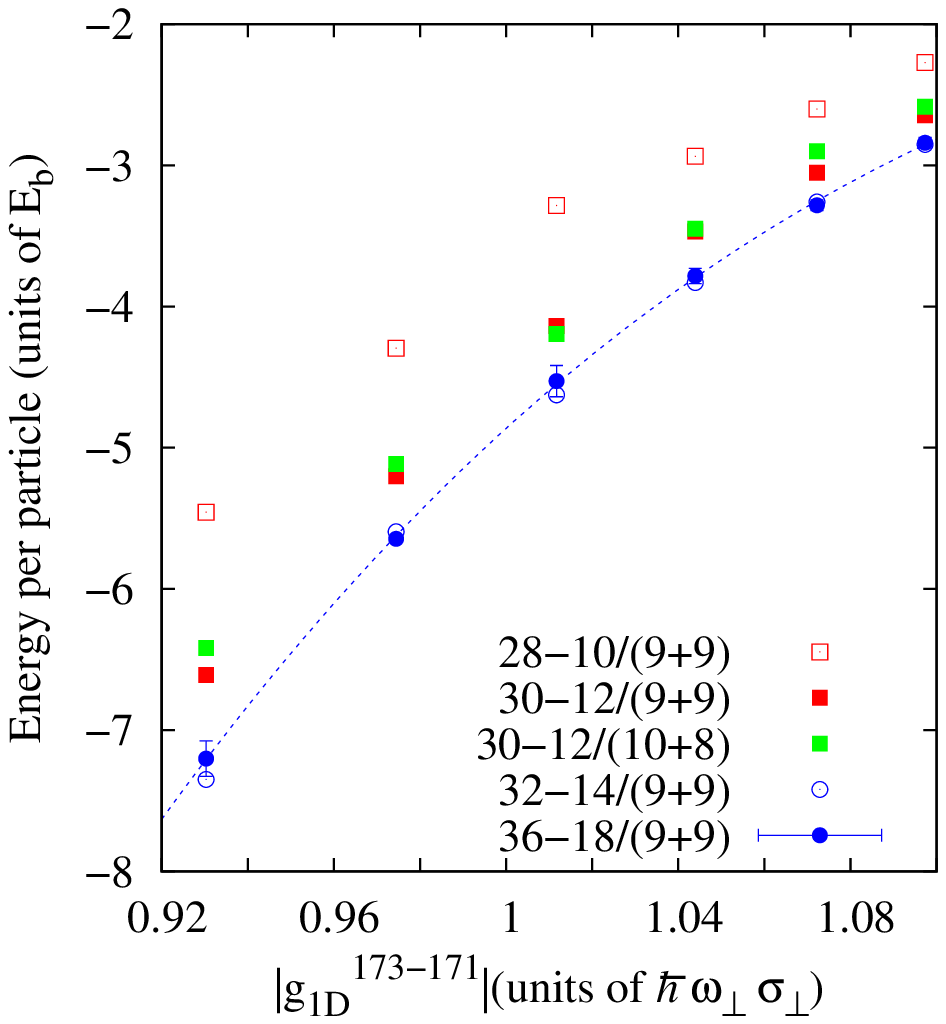}
\caption{Energies per Yb atom for clusters of different compositions (see text for definitions) as a function 
of the interaction parameter between atoms of different 
isotopes. Error bars in all cases are similar to the ones
displayed and not shown for simplicity. Dotted line is a least-squares third orden polynomial fit to the 36-18/(9+9) case and it is intended as a guide-to-the eye.
}
\label{fig1}
\end{center}
\end{figure}

\begin{figure}
\begin{center}
\includegraphics[width=0.8\linewidth]{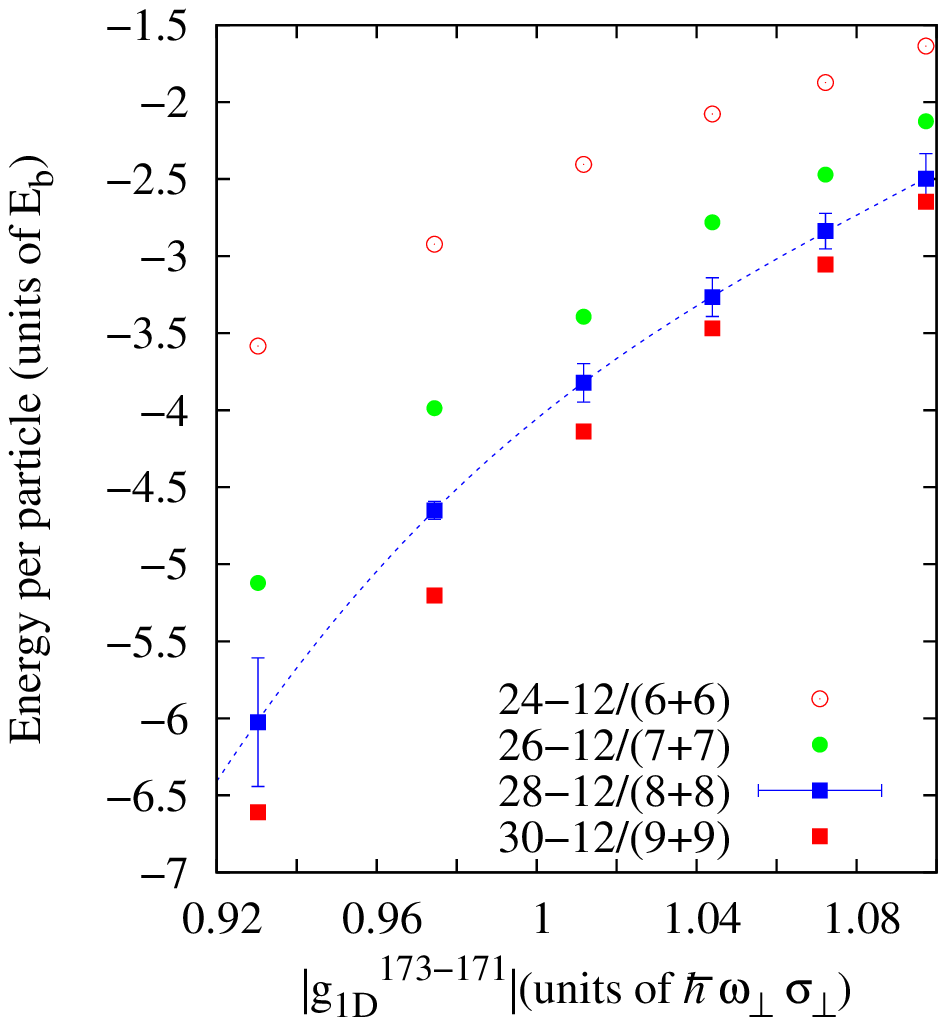}
\caption{Same as in Fig. \ref{fig1} but for clusters in which the number of atoms 
in the spin-polarized $^{173}$Yb subcluster is fixed. As in the previous figure, the error bars are comparable in all cases and only shown for the 28-12/(8+8) droplet for clarity. The dotted line is also intended as a guide-to-the-eye. 
}
\label{fig2}
\end{center}
\end{figure}

\begin{figure}
\begin{center}
\includegraphics[width=0.8\linewidth]{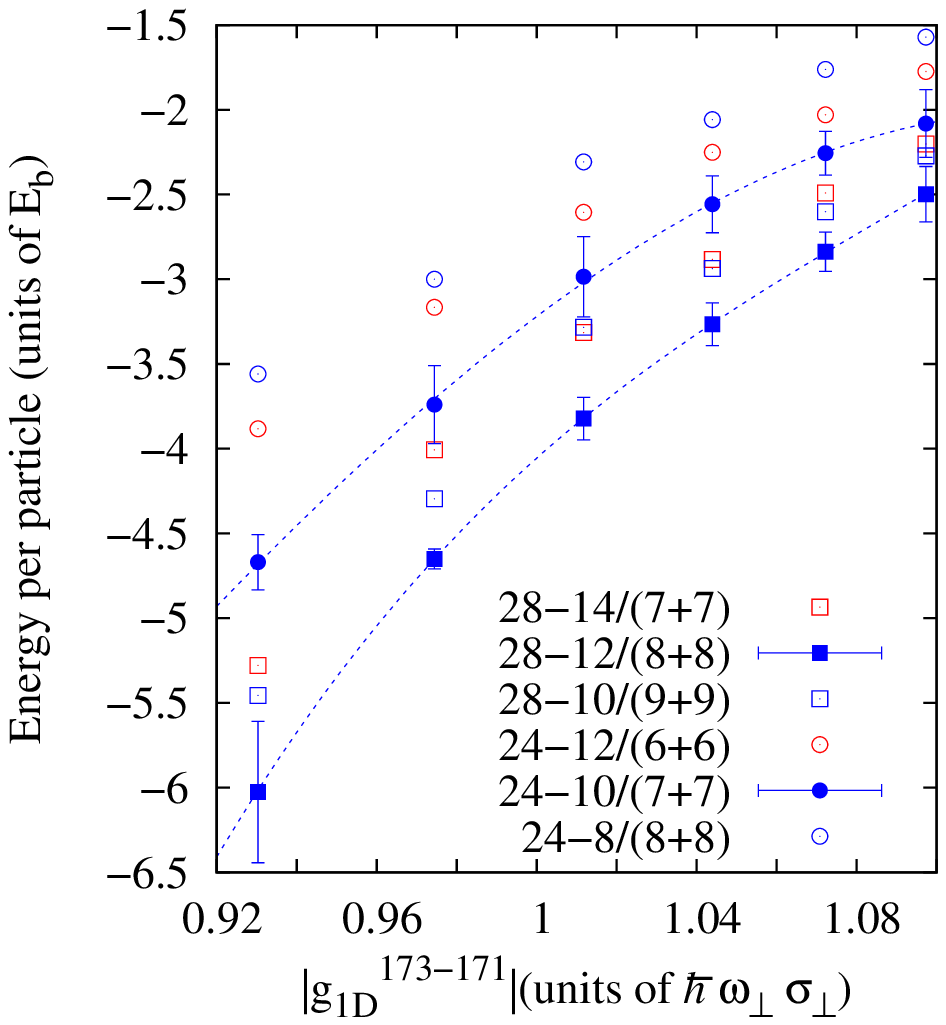}
\caption{Energies per particle for different cluster compositions.  In all cases the error bars are of the size of the ones displayed and the dotted lines are again guides-to-the eye.
}
\label{fig3}
\end{center}
\end{figure}

In Figs. \ref{fig1}-\ref{fig3} and Fig. \ref{fig4},  we display the dependence of the energy per particle 
on the cluster composition to check whether is smaller than $(N_{173}/N_p) E_b$ 
(or $(N_{171}/N_p) E_b$, depending on the case)  and 
self-bound droplets are possible.  The $|g_{1D}^{173,171}|$ range displayed is the result of 
taking $\omega_{\perp}$'s in the interval $2 \pi \times$50-100 Hz, deducing
the corresponding $\sigma_{\perp}$'s, and introducing those values in
Eq. \ref{a1D} to produce the magnitude of the interaction.   
The first of those figures show the case for clusters with total number of particles in the range 24-36, assembled by joining together a variable number of 
spin-polarized $^{173}$Yb atoms and a evenly distributed set of spin-up and spin-down $^{171}$Yb particles.  Since $^{171}$Yb atoms have SU(2) symmetry,  we cannot consider more than two spin sets.  To consider larger clusters in order 
to extrapolate our DMC results to the thermodynamic limit $N_p \rightarrow \infty$  is computationally very expensive and beyond the scope of this work.  
In this example, we fixed $N_{171}$/2 = 9, but the results are similar for clusters with different compositions. To name those clusters, we used the convention $N_p$-$N_{173}$/($N_{171}$/2+$N_{171}$/2). Since their energies per particle are smaller than $E_b$, we may say that, in principle,  the droplets are self-bound, this being due to the attractive interactions between atoms in different (with unequal spins for the atoms in the $^{171}$Yb isotope) molecules. We can see also that, for the same total number of $^{171}$Yb atoms, the stability of the cluster increases with the number of $^{173}$Yb  
particles.  In Fig. \ref{fig1} we display also the case of two unequal sets of spin-up and spin-down $^{171}$Yb atoms (30-12/(10+8)).  The energy per particle is virtually identical that that of the more balanced 30-12/(9+9) one.  This implies that our results are robust with respect to a slight change in spin composition.  Larger differences in the number of spin up and spin down $^{171}$Yb numbers produce unstable clusters (see below). On the other hand, when 
we keep $N_{173}$ constant,  the energy per particle also decreases with $N_{171}$,  as can be  seen in Fig. \ref{fig2}.  Last, we display in 
Fig. \ref{fig3},  a couple of representative examples that indicate that, when $N_p$ is constant, the most stable arrangements  are those with a small imbalance in the $N_{173}/N_{171}$ ratio. 

\begin{figure}
\begin{center}
\includegraphics[width=0.8\linewidth]{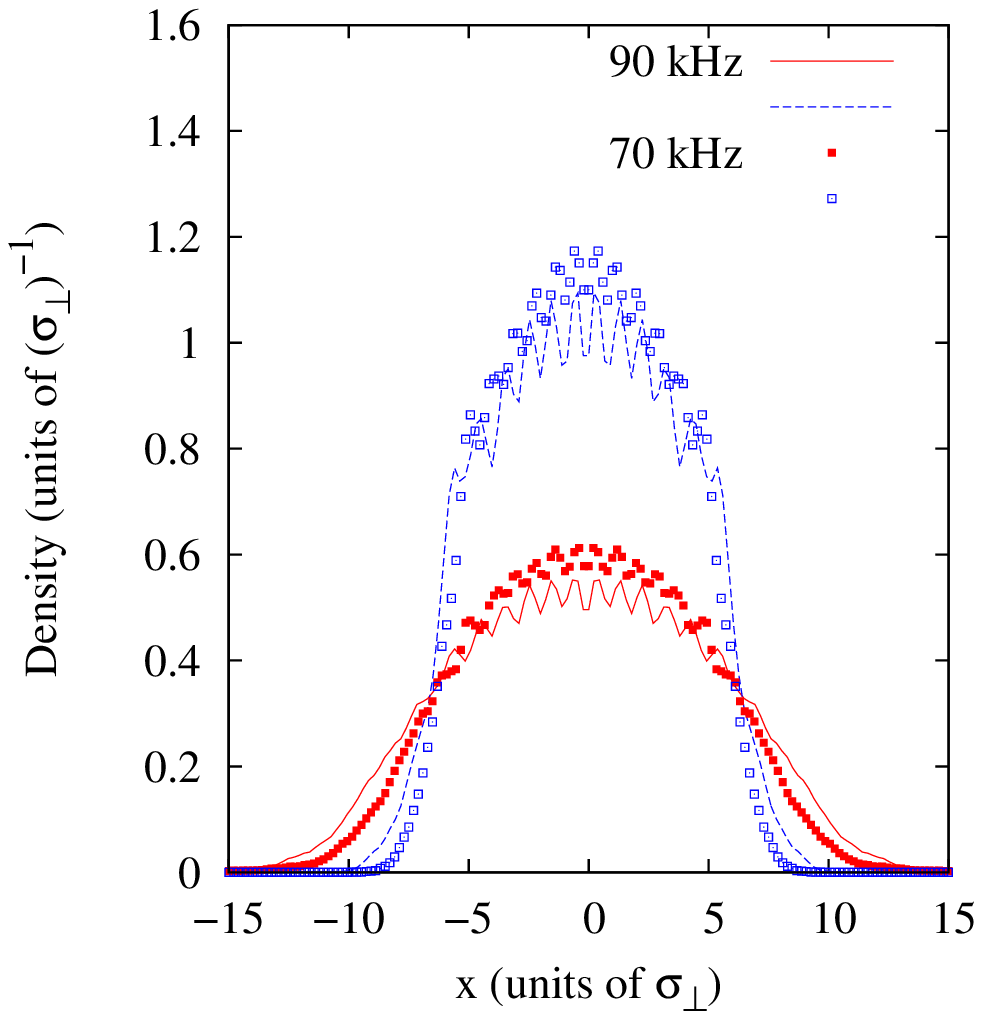}
\caption{Density profiles for a 28-12/(8+8) cluster for two different values of transverse confinement. Open symbols and dashed line correspond to $^{173}$Yb
densities, while solid symbols and full line display the behavior of each spin 
component of the $^{171}$Yb isotope. Error bars are of the size of the symbols and not displayed by simplicity. 
}
\label{perfil1}
\end{center}
\end{figure}

However, to make sure that we have a stable self-bound droplet we have to check that
the cluster will not eventually break during the course of the simulation. To do that, we calculated the density profiles for different arrangements and checked that they remained invariant and finite in width throughout each simulation. As a example of such circumstance, we display in Fig. \ref{perfil1} one of those profiles for a 28-12/(8+8) droplet, representative of clusters
in which the spin-polarized component is $^{173}$Yb, for two different values of transverse confinement.
The profiles are normalized to the total number of particles for the $^{173}$Yb isotope (12) and to the number of particles per spin for the $^{171}$Yb one (8). Those are equilibrium profiles, unchanged along a DMC simulation comprising 3 $\times$ 10$^5$ Monte Carlo steps after thermalization and averaged over 3 independent Monte Carlo histories. To avoid spurious correlations, we considered only configurations separated 100 steps apart, i.e., 
we kept 3000 sets of data. To be sure about stability of the clusters, we compared those total averages with the ones obtained considering the first 1000 DMC steps, the 1000 in the middle, and the 1000 final configurations of each history. In all cases, the results were identical to those shown in Fig. \ref{perfil1}.  As to the properties of the droplet itself, we can see that the tighter the confinement, the larger the width of the atoms of the cluster in the longitudinal direction, in accordance with what happens in balanced clusters \cite{yoprr}. 
We can see also that the $^{171}$Yb densities spread outside the locations of the spin-polarized $^{173}$Yb isotope,  but not too far.  This allows the excess $^{171}$Yb atoms in the wings to bind with the $^{173}$Yb's closer to them.   
 
\begin{figure}
\begin{center}
\includegraphics[width=0.8\linewidth]{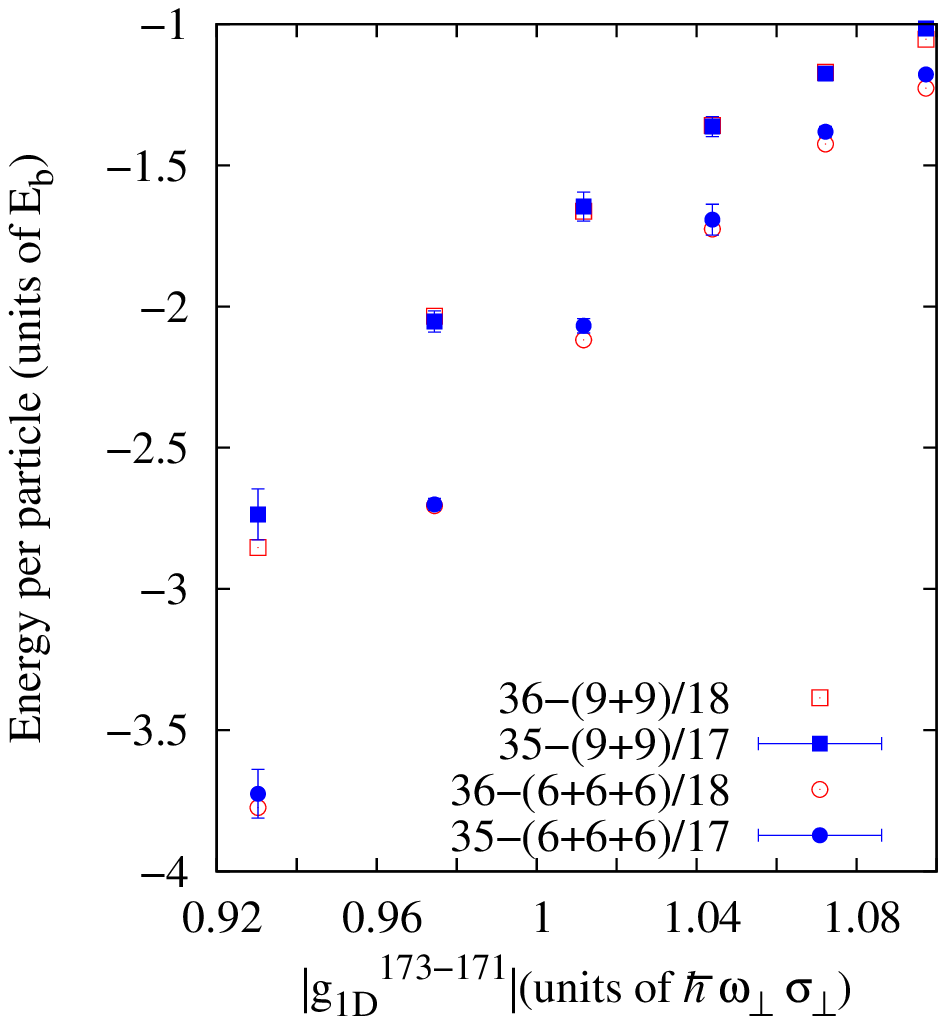}
\caption{Energies per particle for stable clusters in which the spin-polarized component is $^{171}$Yb. Error bars are similar in both cases, but they are only shown for the 35-(9+9)/17 and 35-(6+6+6)/17 clusters. 
}
\label{fig4}
\end{center}
\end{figure}

\begin{figure}
\begin{center}
\includegraphics[width=0.8\linewidth]{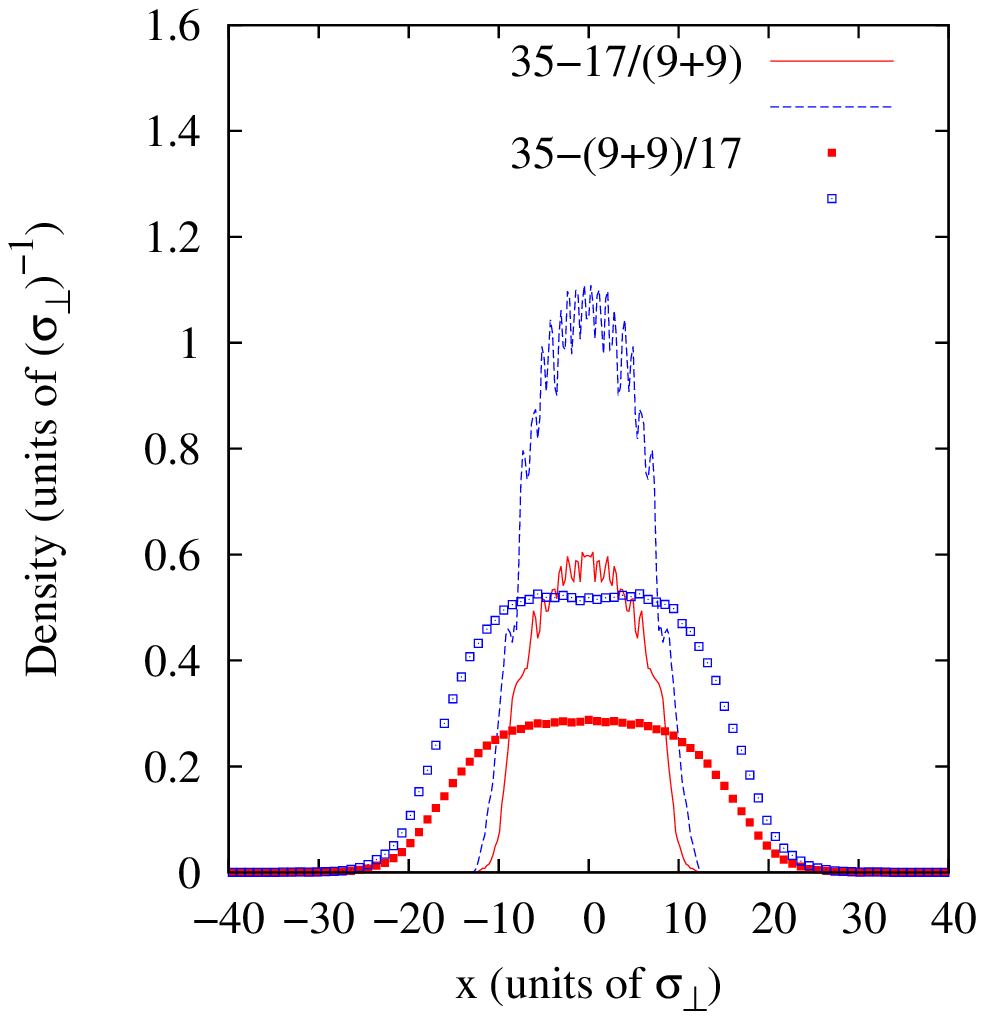}
\caption{Same as in Fig. \ref{perfil1} for two different droplets with the same size and different compositions for a transverse confinement of $\omega_{\perp}$ = 2$\pi \times$90 Hz. Dashed lines and open symbols correspond to the spin-polarized component, while solid lines and symbols show the results for one of the spin components corresponding to atoms in the other isotope. 
}
\label{perfil2}
\end{center}
\end{figure}

Fig. \ref{fig4} gives us the same information as Fig. \ref{fig1}, but for arrangements in which the spin-polarized component belongs to the $^{171}$Yb isotope. This implies that the interactions between the two (or three) sets of $^{173}$Yb atoms are repulsive, as corresponds to the positive three-dimensional scattering length \cite{pra77} between them. 
In that figure, we only display energies per particle for four clusters.  The reason is that they are the only ones for which the density profiles are stable according to the criterion described in the previous paragraph. When the number of $^{171}$Yb decreases (or, equivalently, the number of $^{173}$Yb increases), the clusters end up either splitting into smaller units or regularly spreading along the simulation runs with no equilibrium final position. This last circumstance would be akin to "evaporation".  This happens also when we considered clusters in 
which the spin-polarized part corresponds to $^{173}$Yb and the two-spin $^{171}$Yb sets are too unequal,  for instance for a 30-12/(12+6) arrangement or for the case in which the number of unequal sets of spins is three (out of the 6 possible for the $^{173}$Yb SU(6) atoms), as in the 35-(6+6+6)/17 atoms.
The density profile for the stable 35-(9+9)/17 cluster is displayed in Fig. \ref{perfil2}. There, we can see that this droplet is wider than its counterpart of the same size. Obviously, this is due to the repulsive interactions between atoms in the (9+9) subcluster. In any case, the nine atom subunits are contained within the limits of their spin-polarized counterparts and the whole cluster is stable due to the attractive $^{171}$Yb-$^{173}$Yb interactions.  

\section{Conclusions}

In this work we have dealt with the possibility of having 1D self-bound unbalanced clusters of fermions.  By that, we meant mixtures of Ytterbium isotopes in which the total number of $^{173}$Yb atoms is different than the total number of $^{171}$Yb particles.  As in the case of balanced droplets ($N_{173}$=$N_{171}$),  when both 
components are spin-polarized,  it is impossible to have a self-bound system.  On the other hand, when the atoms 
of one of the isotopes have different spin values,  those droplets are stable.  In this work we have considered mainly examples of evenly split spin populations, but the results are similar for other distributions. 

We saw also that  the relative stability of the droplets depends on their composition.  Unbalanced droplets in which the interactions between particles of the same isotope and unequal spins are repulsive have a very narrow stability range.  What we have found is that when $N_{173}$ (two or three different spins) $>$ ($N_{171}$+1) the clusters either break into smaller units or evaporate.  
Conversely, when the unequal-spin atoms attract each other (spin-up and spin-down $^{171}$Yb),  the variability in the cluster compositions is larger. In particular, for the majority of the cases considered in this work, i.e., 
for $N_p$-$N_{173}$/($N_{171}$/2+$N_{171}$/2) arrangements
with fixed $N_p$,  we found that the
most stable droplets were those of the type $N_p$-($N_p/2$-2)/($N_p/4$+1,$N_p/4$+1), 
as can be seen in Fig. \ref{fig3}.  That can be 
understood as the result of having a  
balanced cluster with two additional $^{171}$Yb atoms located in both wings, as can be seen in the density profiles of the 28-12/(8+8) arrangement displayed in Fig.  \ref{perfil1}. 
The relative reduction in the number of spin-polarized $^{173}$Yb atoms decreases the effective repulsive interaction between identical fermions,  lowering the total energy per particle.  
However, if we further deplete the $^{173}$Yb part of the cluster, the reduction in the $^{171}$Yb-$^{173}$Yb interactions
de-stabilize the entire structure. This is because the $^{171}$Yb atoms in the outer part of the wings are progressively  further away from the $^{173}$Yb's at the center, and unable to form molecules.  In addition,  
those $^{171}$Yb atoms should bind to other $^{171}$Yb's to stay in
the cluster,  the $^{171}$Yb-$^{171}$Yb interactions being too weak 
to stabilize the wings if they are made up of more than two $^{171}$Yb atoms.

\begin{acknowledgments}
We acknowledge financial support from Ministerio de
Ciencia e Innovación MCIN/AEI/10.13039/501100011033 
(Spain) under Grant No. PID2020-113565GB-C22
and from Junta de Andaluc\'{\i}a group PAIDI-205.  
We also acknowledge the use of the C3UPO computer facilities at the Universidad
Pablo de Olavide.
\end{acknowledgments}

\bibliography{unbalanced}

\end{document}